 \documentclass[aps,prd,10pt,showpacs,amsmath,twocolumn,floatfix,amssymb,nofootinbib,longbibliography]{revtex4-2}
\usepackage{graphicx}
\usepackage{comment}
\usepackage[usenames]{color}
\usepackage{bm}
\usepackage{ifpdf}
\usepackage{floatrow}
\usepackage{makecell}
\usepackage{caption}
\usepackage{stackrel}
\usepackage{subcaption}
\usepackage{enumitem}

\usepackage{caption}
\usepackage[normalem]{ulem}
\usepackage[dvipsnames]{xcolor}
\usepackage[utf8]{inputenc}
\usepackage{hyperref}
\hypersetup{
  pdfnewwindow=true,      
  colorlinks=true,        
  linkcolor=PineGreen,    
  citecolor=PineGreen,    
  filecolor=PineGreen,    
  urlcolor=PineGreen      
}
\newcommand{\be}{\begin{eqnarray}}
\newcommand{\ee}{\end{eqnarray}}
\newcommand{\non}{\nonumber \\}

\begin{document}

\title{Scattering phase shift in quantum mechanics on quantum computers: \\non-Hermitian systems and imaginary-time simulations}

\author{Peng~Guo}
\email{peng.guo@dsu.edu}
\affiliation{College of Arts and Sciences,  Dakota State University, Madison, SD 57042, USA}

\author{Paul~LeVan}
\email{paul.levan@dsu.edu}
 \affiliation{College of Arts and Sciences,  Dakota State University, Madison, SD 57042, USA}

\author{Frank~X.~Lee}
\email{fxlee@gwu.edu}
\affiliation{Department of Physics, George Washington University, Washington, DC 20052, USA}

\author{Yong~Zhao}
\email{yong.zhao@anl.gov}
\affiliation{Physics Division, Argonne National Laboratory, Lemont, Illinois 60439, USA}

\date{\today}

\begin{abstract}
To overcome the fast oscillatory behavior of correlation functions for extracting scattering phase shift in real-time quantum simulations encountered in Ref.\cite{Guo:2026qkx}, we propose and test two solutions in the present work. One is to simulate Hermitian systems in imaginary time, the other is to simulate non-Hermitian systems in real time. We demonstrate that both approaches lead to the problem of non-unitary quantum evolution which can be solved by combining two quantum algorithms: block encoding and Hadamard test. The combined quantum algorithm does not require mid-circuit measurements or adjustment of the input parameters of the Hamiltonian, and can be easily implemented on quantum computers.
Numerical tests on quantum simulators show that both approaches agree with exact solutions for a sufficiently long time before the signal is lost in statistical fluctuations. The results bode well for using non-Hermitian and imaginary-time simulations to circumvent oscillations inherent in real-time simulation of other quantum systems.  In particular, the non-Hermitian  approach shows  a decisive advantage over the imaginary-time one  on the number of required ancillary qubits, and hence is more practical to scale up.  
\end{abstract}

\maketitle

\section{Introduction}\label{sec:intro}

Scattering is an indispensable tool in our understanding of interactions in nature, 
from the original Rutherford experiment on the structure of the atom to modern experiments in nuclear and particle physics.

 The promise of quantum computing to solve scattering problems is drawing increasing attention.
 Various ideas have been proposed in nuclear physics and lattice QCD in general~\cite{itou2025,Yamamoto_2024,Zhang_2021,Alexandru_2024}; and scattering in particular~\cite{davoudi2025,farrell2025,ingoldby2025,schuhmacher2025,Wang_2024,Li_2024,yusf2024,Sharma_2024,Brice_o_2021}.  
Additionally,
there have been developments on the subject of reaction processes in quantum simulations, such as combined  Variational Quantum Eigensolver    and   L\"uscher-  or Busch-Englert-Rza\.zewski-Wilkens (BERW)-like formula approach in Ref.~\cite{PhysRevC.109.064623,PhysRevC.110.054604};  measuring phase shift by wave packet time delay \cite{PhysRevD.104.054507};  reconstructing scattering amplitudes through the coupling of the particle with an ancillary spin-1/2 \cite{Mussardo2024}; radiative processes \cite{Bedaque:2022ftd};   and other ideas, e.g. \cite{Gustafson:2020yfe,Briceno:2020rar}.

In a previous study~\cite{Guo:2026qkx}, we explored the feasibility of extracting scattering phase shift on quantum computers in 1D quantum mechanics using the integrated correlation function (ICF) formalism~\cite{Guo:2023ecc},  and was faced with the challenge of fast oscillatory behavior typically encountered in real-time quantum simulations. Different ideas have been proposed to mitigate the issue by post-data analysis, see e.g. \cite{Guo:2025vgk,Burbano:2025pef}. We explored two such methods, one is $E\to E+i\epsilon$ prescription, the other $L \rightarrow i L$ rotation,  where $E$ refers the energy of a quantum system and $L$ is the size of periodic box for a trapped system. We showed that both methods allow the extraction of infinite volume phase shifts, with the $L \rightarrow i L$ rotation   being more effective than the $E\to E+i\epsilon$ prescription. Additional post-data processing approaches include averaging   either in time or energy bins~\cite{Guo:2025vgk,Burbano:2025pef}.  However, $ iL$ rotation yields a non-Hermitian Hamiltonian, the question of how to deal with quantum simulation of a non-Hermitian Hamiltonian system on unitary-gate based quantum hardware was left unanswered.

The goal of the present work is to expand   the discussion of our previous work in Ref.\cite{Guo:2026qkx} to explore the real time quantum simulation    of    non-Hermitian systems.  The same quantum algorithm   can be applied to imaginary time evolution of a quantum system as well, hence      the imaginary time quantum simulation is also studied in this work.
We will employ the same 1D quantum mechanical setup with the same contact interaction as in Ref.~\cite{Guo:2026qkx}. 
We consider two possibilities. One is to rotate time and simulate the system in imaginary time. If the Hamiltonian is independent of time, it stays Hermitian.  This approach is the one widely adopted in Monte Carlo simulations in lattice QCD and other quantum field theories. Except here we are not performing Monte Carlo, but imaginary time quantum  evolution.
The other possibility is to rotate space and simulate the system in real time.
The  rotation in space $L \rightarrow i L$ renders the system Hamiltonian non-Hermitian.
In both cases, we are left with the challenge of how to devise quantum algorithms and circuits to carry out time evolution of  non-unitary operators.   We will show that both imaginary time quantum simulation of the integrated correlation function   and real time quantum simulation of   a non-Hermitian system  can be realized by using Hadamard test algorithm \cite{PhysRevA.75.012328} combined with block encoding method used for imaginary time evolution in Refs.~\cite{PhysRevA.109.052414,Leadbeater_2024,Yi:2025zpa}.

The paper is organized as follows. 
In Sec.~\ref{sec:scatt1DQM},  the integrated correlation function formalism in 1D quantum mechanical scattering is outlined.  In Sec.~\ref{sec:algorithm}, the quantum algorithm for trace of non-unitary operators is presented along with their quantum circuits.  
In Sec.~\ref{sec:numerics}, numerical tests  on quantum simulators are discussed.  The optimization of quantum circuits for real time non-Hermitian quantum simulation is given in Sec.~\ref{sec:optimization},  followed by summary and outlook in Sec.~\ref{sec:summary}.

\section{Summary of integrated correlation function approach to 1D Quantum Mechanical scattering}\label{sec:scatt1DQM}

In this section, we give a brief outline of the essential elements in the integrated correlation function approach to the scattering problem in 1D quantum mechanics (QM). The complete discussion can be found in Ref.~\cite{Guo:2026qkx}.

\subsection{Hamiltonian matrix}
As discussed in \cite{Guo:2023ecc,Guo:2024zal,Guo:2024pvt,Guo:2025ngh,Guo:2025vgk,Guo:2025lmd,Guo:2026qkx},  the infinite volume  elastic scattering phase shift, $\delta (\epsilon )$, is   related  to the difference of integrated   correlation functions in a trap through a weighted integral,
\begin{equation}
\triangle C(\tau)=C(t) - C_0 (t) \stackrel{\mbox{trap} \rightarrow \infty}{\rightarrow} \frac{i t}{\pi} \int_0^{\infty} d \epsilon \delta(\epsilon) e^{- i \epsilon t},  \label{ICFmainEQ}
\end{equation}
 where $C(t)$ and $C_0 (t)$ are the integrated correlation functions (ICF) for   interacting and non-interacting  particles in the trap.   In quantum mechanics, the integrated correlation function is associated with the Hamiltonian of non-relativistic two-particle system in the trap,
 \begin{equation}
 C(t) = \mbox{Tr} \left [ e^{- i \hat{H} t} \right ]  .
 \end{equation}
The same holds for non-interacting integrated correlation function $C_0 (t)$.  

The discretized Hamiltonian matrix for a particle interacting with a contact interaction potential placed at center of a periodic box is written as~\cite{Guo:2026qkx},
\begin{align}
\hat{H} & = - \frac{1}{2 m a^2}  \sum_{\alpha = 0}^{2^\Gamma -1}     \left ( | \alpha \rangle \langle \alpha +1 |  + | \alpha +1 \rangle \langle \alpha  |  \right )   \nonumber \\
 & +  \sum_{\alpha = 0}^{2^\Gamma -1}  \left ( \frac{1}{ m a^2}  + V(x_\alpha ) \right )    | \alpha \rangle \langle \alpha  |      , \label{Hcoordinate}
\end{align}
where $\Gamma$ denotes the number of qubits, $m$ is the mass of particle and $| \alpha \rangle$ is short-hand notation of discretized position $| x_\alpha \rangle$ basis. The position  $x \in [- \frac{L}{2}, \frac{L}{2}]$  in a periodic box of length $L$ is discretized to  $x_\alpha = - \frac{L}{2} + a \alpha $, where $a = \frac{L}{2^\Gamma}$ is lattice spacing and $\alpha \in [0, 1, \cdots, 2^\Gamma-1]$. The periodic boundary condition requires $ | 2^\Gamma  \rangle = | 0 \rangle $. The   contact interaction potential is defined by,
\begin{align}
V(x_\alpha) & = \begin{cases}  \frac{V_0}{2 a }, & \alpha  \in [ \frac{2^\Gamma}{2}-1 ,  \frac{2^\Gamma}{2} ] \\ 0, & \text{otherwise}  \end{cases}  \nonumber \\
& \xrightarrow[\text{fixed} \  L]{ \substack{ a \rightarrow 0  \\ 2^\Gamma \rightarrow \infty  }} V_0 \delta(x).
\end{align}

The analytic solution of infinite volume scattering phase shift for the contact potential is given by
\begin{equation}
\delta(E) = \cot^{-1} \left ( - \frac{\sqrt{2 m E} }{m V_0} \right ),
\end{equation}  
which leads to the analytic expression of right-hand side of Eq.(\ref{ICFmainEQ}),
 \begin{equation}
\triangle C(t) \stackrel{\mbox{trap} \rightarrow \infty}{\rightarrow}  \frac{1}{2} \mbox{erfc}  \left( m V_0  \sqrt{\frac{i t}{2m}}\right) e^{ (m V_0)^2 \frac{i t}{2m}} - \frac{1}{2}  .  \label{ICFmainEQanalytic}
\end{equation}
It is a fast oscillatory, complex-valued function in real time. 
Its version in imaginary time can be obtained by an analytic continuation $t\to -i\tau$, which causes $\triangle C(\tau)$ to be positive if the interaction is attractive ($V_0<0$), negative if repulsive ($V_0>0$) (see Ref.\cite{Guo:2026qkx}).

\subsection{Quantum circuits for the Hamiltonian}\label{sec:QCHamiltonian}

The quantum circuits for the Hamiltonian matrix in Eq.(\ref{Hcoordinate}) are given in Sec.~III in Ref.~\cite{Guo:2026qkx}, also see e.g. Ref.~\cite{Guo:2025xpd}. For the purposes of this work, we need to express the entire Hamiltonian matrix in Pauli basis.
Starting with the even-odd decomposition plus interaction:
\begin{align}
& \hat{H} = \hat{H}_a +\hat{H}_b + \hat{H}_v, \text{ with} \nonumber \\
& \hat{H}_a  = - \frac{1}{2 m a^2}  \sum_{\alpha = 0}^{2^{\Gamma - 1}-1}     \Bigl (\, | 2 \alpha \rangle \langle 2 \alpha +1 |  + | 2 \alpha +1 \rangle \langle 2 \alpha  |  \,\Bigr ) , \nonumber \\
& \hat{H}_b   \nonumber \\
&  = - \frac{1}{2 m a^2}  \sum_{\alpha = 0}^{2^{\Gamma - 1}-1}     \Bigl (\, | 2 \alpha  + 1 \rangle \langle 2 \alpha +2 |  + | 2 \alpha +2 \rangle \langle 2 \alpha  + 1 |  \,\Bigr ) , \nonumber \\
& \hat{H}_v   =  \sum_{\alpha = 0}^{2^{\Gamma}-1}  \left ( \frac{1}{ m a^2}  + V(x_\alpha ) \right )    | \alpha \rangle \langle \alpha  |  . \label{HaHbHvcoordinate}
\end{align}

For \(\hat{H_a}\), the Pauli expansion  is simply the tensor product of X gate and identity gates: \(I_{\Gamma} \otimes\ldots\otimes I_2\otimes X_1\).  
For \(\hat{H}_v\), it was worked out in Eq.(33) of Ref.~\cite{Guo:2026qkx}.

For \(\hat{H}_b\), the Pauli expansion is more complicated, with expression:
\begin{equation}
\begin{split}
\hat{H}_b
=-&\frac{1}{2ma^2}\Biggl[
\sum_{k=2}^{\Gamma}\frac{1}{2^{k-2}}\\
&\sum_{S}
(-1)^{\frac{|S|}{2}}\,
I^{\otimes (\Gamma-k)}\otimes (X_k-I_k)\otimes P_{k-1}(S)\Biggr]
\end{split}
\label{hatB_main}
\end{equation}
where the inner sum is over all even sized subsets of \(\{1, \ldots, k-1\}\) and
\begin{equation}
\begin{split}
&P_{k-1}(S)=M_{k-1}(S)\otimes\cdots\otimes M_1(S)\\
&M_j(S)=
\begin{cases}
Y_j, & j\in S,\\
X_j, & j\notin S.
\end{cases}
\end{split}
\label{hatB_auxs}
\end{equation}
The derivation is given in Appendix~\ref{sec:appendix-A}.

In general, the Pauli expansion is a lengthy and tedious process, but for few-qubit systems,  the Hamiltonian matrix takes fairly simple forms.
For one qubit, it is given by $X$ and $I$ gates,
\begin{equation}
\hat{H} = - \frac{1}{m a^2} X + \left ( \frac{1}{m a^2} + \frac{V_0}{2 a} \right ) I,
\label{Honequbit}
\end{equation}
and for two qubits by $X$, $I$, and $Z$ gates,
\begin{align}
\hat{H}  &=  - \frac{1}{2 m a^2} \left [ I_2 \otimes  X_1 + X_2 \otimes    X_1 \right ]   -  \frac{V_0}{4 a}    Z_2 \otimes Z_1  \nonumber \\
& +  \left (  \frac{1}{m a^2} + \frac{V_0}{4 a} \right )  I_2 \otimes I_1.
\label{Htwoqubit}
\end{align}

Under the $L \rightarrow iL$ rotation, the Hamiltonian matrix becomes  non-Hermitian, 
\begin{equation}
\hat{H}^{(iL)} =  \frac{1}{m a^2} X - \left ( \frac{1}{m a^2} +  i \frac{V_0}{2 a} \right ) I, \label{HiLonequbit}
\end{equation}
for one qubit, and 
\begin{align}
\hat{H}^{(iL)}  &=   \frac{1}{2 m a^2} \left [ I_2 \otimes  X_1 + X_2 \otimes    X_1 \right ]   + i   \frac{V_0}{4 a}    Z_2 \otimes Z_1  \nonumber \\
& -  \left (  \frac{1}{m a^2} + i \frac{V_0}{4 a} \right )  I_2 \otimes I_1  , \label{HiLtwoqubit}
\end{align}
for two qubits.
The non-Hermitian Hamiltonian has the general form,
\begin{equation}
    \hat{H}^{(iL)} = H_1 + i H_2,
    \label{H1H2}
\end{equation}
where $H_1$ is Hermitian and $iH_2$ is anti-Hermitian. Both $H_1$ and $H_2$ are Hermitian matrices.

\section{Quantum algorithm for trace of non-unitary operators}\label{sec:algorithm} 

Symbolically, we can suppress the real-time, oscillatory operator $e^{-iHt}$ in two ways. One is by rotating to imaginary time $t\to -i \tau$, so $e^{-iHt} \to e^{-H\tau} $. The Hermitian Hamiltonian is untouched.
The other is by rotating space $L\to i L$, so $e^{-iHt} \to e^{-iH^{(iL)}t} \to e^{-iH_1t} e^{H_2 t}$.
In both cases, we must deal with the problem of a non-unitary operator: $e^{-H\tau}$ in imaginary-time evolution, or $ e^{H_2 t}$ in real-time evolution. Whether it is an exponential growth or decay depends on whether it is a bound or a scattering state in the system.

\subsection{Trace of non-unitary operator}
 
The problem boils down to taking the trace of a non-unitary operator  $A$,
\begin{equation}
\mbox{Tr} \left [A\right ] = \sum_{\alpha}  \langle \alpha |A | \alpha \rangle,
\end{equation}
which  can be accomplished by combining two quantum algorithms~\cite{PhysRevA.75.012328}: block encoding and Hadamard test.

 First, the non-unitary operator $A$ has to be block encoded in a unitary matrix $U_A$ with $n$ ancillary qubits,
\begin{equation}
U_A = \begin{pmatrix} A & \cdot \\ \cdot  & \cdot \end{pmatrix} ,
\end{equation}
which can then be projected to the non-unitary sector,
\begin{equation}
U_A \left ( | \alpha \rangle \otimes | 0 \rangle^{\otimes n} \right ) =  A | \alpha \rangle  \otimes | 0 \rangle^{\otimes n}  + \cdots, \label{UAblockencode}
\end{equation}
discarding everything else from the non-A elements in the $U_A$ matrix as represented by the ellipsis.  The implementation of block encoding for real time  evolution of non-Hermitian Hamiltonian systems or imaginary time evolution of  Hermitian systems  will be discussed in Sec.~\ref{sec:blockencode} and Sec.~\ref{sec:blockencodeITE}.

  \begin{figure}[h]
  \frame{\includegraphics[width=0.99\textwidth]{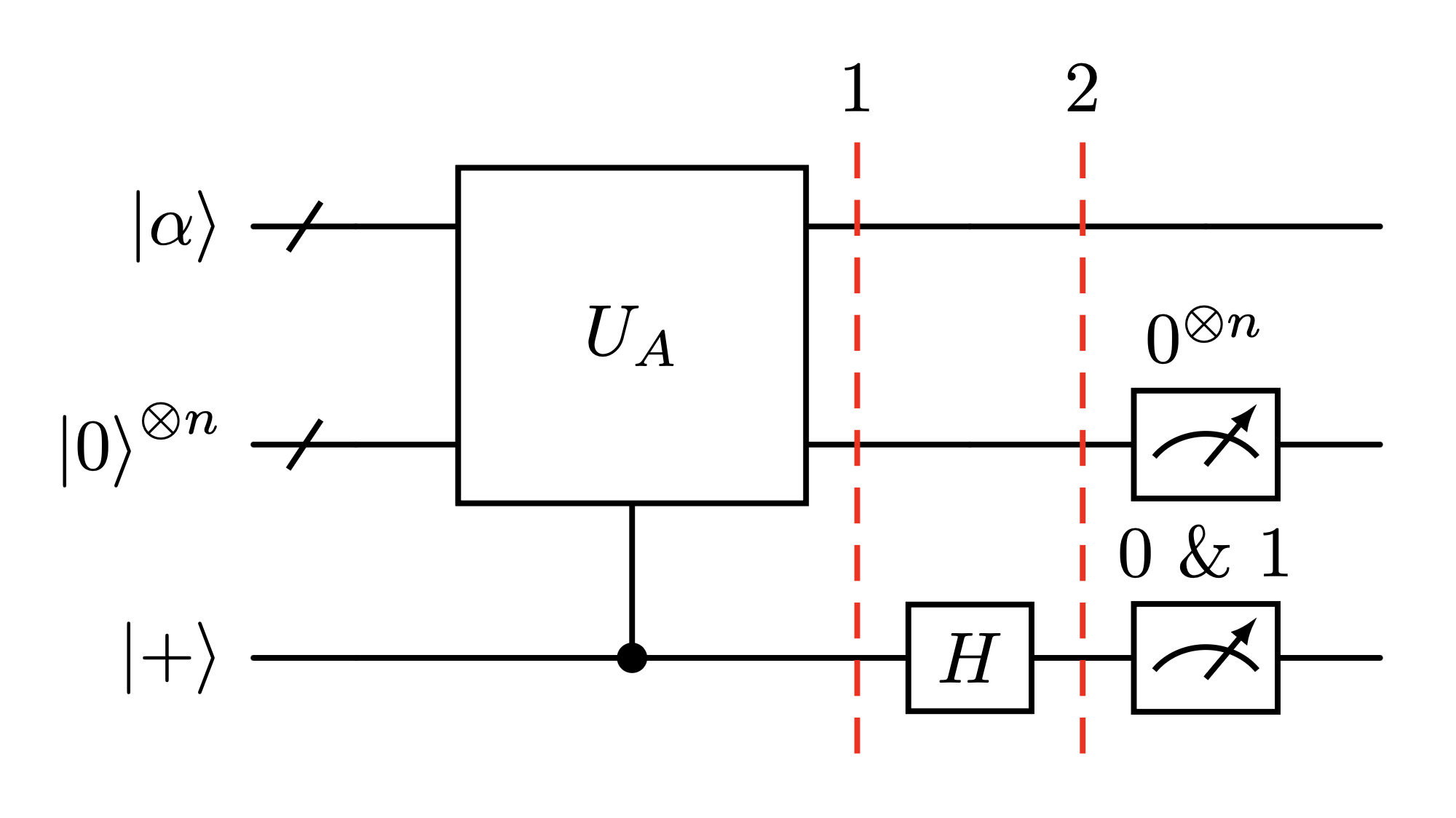}}
\caption{ Quantum circuit for computing real part of $\langle \alpha | \frac{A^\dag + A}{2} | \alpha \rangle$, where non-unitary operator $A$ is block encoded in unitary operator $U$.   $|+\rangle $ refers the linear superposition state $H |0\rangle=\frac{|0\rangle + |1 \rangle}{\sqrt{2}}$.}
\label{blockhadamardQC}
\end{figure} 

Next, combining with Hadamard test algorithm \cite{PhysRevA.75.012328} allows us to evaluate $\mbox{Tr}[A]$. For the real part, starting with quantum circuit in Fig.~\ref{blockhadamardQC}, the quantum state at step-1 is given by,
\begin{equation}
| \psi_1 \rangle  = \frac{1}{\sqrt{2}} | \alpha \rangle  \otimes | 0 \rangle^{\otimes n} \otimes | 0 \rangle +  \frac{1}{\sqrt{2}}  A | \alpha \rangle   \otimes | 0 \rangle^{\otimes n} \otimes | 1 \rangle  + \cdots .
\end{equation}
Applying another Hadamard gate on the last ancillary qubit for Hadamard test algorithm at step-2 then yields,
\begin{equation}
| \psi_2 \rangle  =\frac{1+A }{2}  | \alpha \rangle  \otimes | 0 \rangle^{\otimes n} \otimes | 0 \rangle +  \frac{1-A }{2}  | \alpha \rangle   \otimes | 0 \rangle^{\otimes n} \otimes | 1 \rangle  + \cdots .
\end{equation}
Measurements will be carried out on $n+1$ ancillary qubits (n for block encoding and one for Hadamard test algorithm), only keeping zeros on all $n$ ancillary qubits for block encoding, and both zero (labeled as 0) and one (labeled as 1) for ancillary qubit for the Hadamard test algorithm.
The probability difference is  given by
\begin{equation}
P_\alpha (0) - P_\alpha (1) = \langle \alpha | \frac{A^\dag + A}{2} | \alpha \rangle,
\end{equation}
where $P_\alpha (0) $ and $P_\alpha (1)$ refer to the probabilities of finding all $n$ ancillary qubits at zero states and the ancillary qubit for Hadamard test algorithm at state zero, and state one respectively.

Finally by summing over all the $| \alpha \rangle$ basis states and using $\mbox{Tr}[A^\dag ] = \mbox{Tr}[A^*]$, we find,
\begin{equation}
\sum_\alpha \left [ P_\alpha (0) - P_\alpha (1) \right ] = \mbox{Re} \left (  \mbox{Tr} [ A ] \right ).
\end{equation}
Similarly, by inserting a $S$-gate in front of  the  Hadamard gate in Fig.~\ref{blockhadamardQC}, the imaginary part can be measured by,
\begin{equation}
\sum_\alpha \left [P_\alpha (1) - P_\alpha (0) \right] = \mbox{Tr}\left [ \frac{A - A^\dag  }{2 i}  \right ] = \mbox{Im} \left (\mbox{Tr}[A] \right ).
\end{equation}

\subsection{Block encoding for real time evolution of non-Hermitian  systems}\label{sec:blockencode}
Here we detail how to implement block encoding for the non-Hermitian Hamiltonian in real-time evolution arising from the $L\to iL$ rotation, using a  technique developed in Refs.~\cite{PhysRevA.109.052414,Leadbeater_2024,Yi:2025zpa}. In general, a non-Hermitian Hamiltonian can be split into  two groups (Hermitian and anti-Hermitian) in Pauli strings,
\begin{equation}
\hat{H} = \sum_i c_i \hat{h}_i + i \sum_{j=1}^n c_j \hat{h}_j , \label{Hgroups}
\end{equation}
where all the coefficients, $c_i$, are real, and all the $\hat{h}_i$'s are tensor product of Pauli gates and are Hermitian. We have assumed $n$ terms in the anti-Hermitian operator group, but kept open the number of terms in the Hermitian operator group.

Using the Hamiltonian in Eq.\eqref{Hgroups}, the real-time evolution of a non-Hermitian system with a single step $\delta t$ is given by,
\begin{equation}
e^{- i \hat{H} \delta t}  \simeq \left ( \prod_{i} e^{- i c_i \hat{h}_i  \delta t}  \right ) \left ( \prod_{j=1}^n e^{ c_j \hat{h}_j  \delta t} \right ) ,  \label{expHdt}
\end{equation}
where the  terms in  the first bracket are all unitary operators, but the terms in the second bracket are all non-unitary operators and have to be   block encoded into  unitary operators by using ancillary qubits.

  \begin{figure}[h]
  \frame{\includegraphics[width=0.99\textwidth]{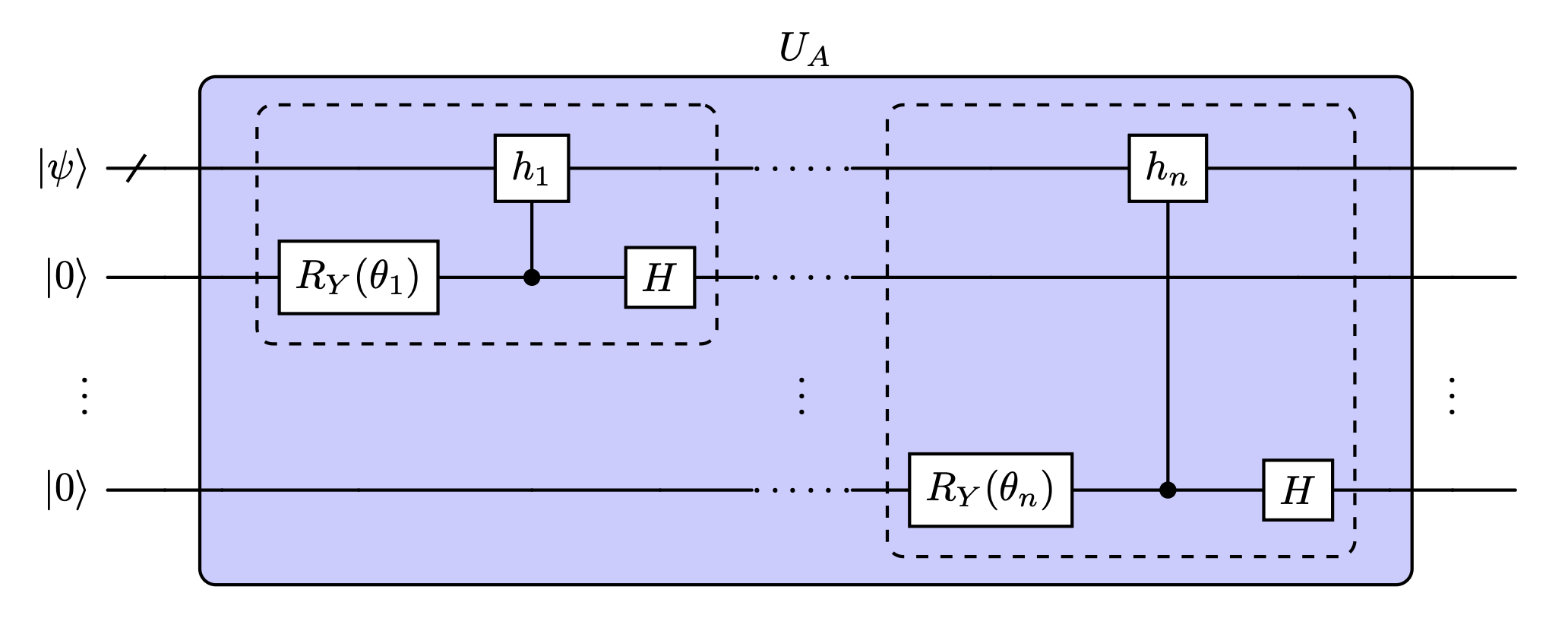}}
\caption{ Block encoding quantum circuit for unitary evolution of non-Hermitian system in real time, see Eq.\eqref{u1} to Eq.\eqref{defryangle}. 
}
\label{ITEonestepQC}
\end{figure}

The quantum circuit  of  block encoding   a single step of time evolution operator, $e^{- i \hat{H} \delta t} $ in Eq.(\ref{expHdt}), into unitary operation, $U_A \left (  | \psi \rangle \otimes | 0 \rangle^{\otimes n} \right ) $, is given in  Fig.~\ref{ITEonestepQC}, where 
\begin{equation}
| \psi \rangle =   \prod_{i} e^{- i c_i \hat{h}_i  \delta t}    | \alpha \rangle ,
\label{u1}
\end{equation}
are pure unitary operations that do not require block encoding.
 The non-unitary operator $A$  in  Fig.~\ref{ITEonestepQC} requires   ancillary qubits for block encoding and is defined  by,
\begin{equation}
A = \prod_{j=1}^n \frac{  e^{ c_j \hat{h}_j  \delta t}  }{\sqrt{2(\alpha_j^2+ \beta_j^2)}} , \label{defA}
\end{equation}
where,
\begin{equation}
 \alpha_j = \cosh \left ( c_j \delta t \right ), \ \ \ \ \beta_j = \sinh \left ( c_j \delta t \right ) . \label{defAalphabeta}
\end{equation} 
Each non-unitary operator for a single time evolution step, $e^{ c_j \hat{h}_j  \delta t} $, requires one ancillary qubit for block encoding  into a unitary operation, and  $n$ non-unitary operators requires $n$ ancillary qubits.
The extra normalization factors $\frac{1}{\sqrt{2(\alpha_j^2+ \beta_j^2)}} $ in Eq.(\ref{defA})   for each   $e^{c_j \hat{h}_j \delta t }  $ term   are generated by the $R_Y (\theta_j)$ gate in block encoding process:
\begin{equation}
R_Y(\theta_j) = \begin{bmatrix}  \frac{\alpha_j}{ \sqrt{\alpha_j^2+ \beta_j^2}} & \frac{\beta_j}{ \sqrt{\alpha_j^2+ \beta_j^2}}  \\ -  \frac{\beta_j}{ \sqrt{\alpha_j^2+ \beta_j^2}}  &  \frac{\alpha_j}{ \sqrt{\alpha_j^2+ \beta_j^2}} \end{bmatrix} ,
\end{equation}
   where the rotation angles  are defined by
\begin{equation}
\theta_j = - 2 \cos^{-1} \left (\frac{\alpha_j}{\sqrt{\alpha_j^2 + \beta_j^2}}  \right ).  \label{defryangle}
\end{equation}

The block encoding method given in Fig.~\ref{ITEonestepQC}  can be implemented for $N$-step time evolution  via the trotterization approximation,
\begin{equation}
e^{- i \hat{H}  t}  \simeq  \left [ \left ( \prod_{i} e^{- i c_i \hat{h}_i  \delta t}  \right ) \left ( \prod_{j=1}^n e^{ c_j \hat{h}_j  \delta t} \right ) \right ]^N,  \;\;t = N \delta t.
\end{equation}
Therefore $n \times N$ ancillary qubits are needed for block encoding of non-unitary operators, and  the number of  required ancillary qubits increases linearly as time increases.

\subsection{Block encoding for imaginary time evolution of  Hermitian systems}\label{sec:blockencodeITE}
As discussed earlier, evolution of Hermitian Hamiltonian in imaginary time involves  $ e^{-H\delta\tau} $, and evolution of non-Hermitian Hamiltonian in real time  involves  $ e^{-iH_1 \delta t} e^{H_2 \delta t}$. They are mathematically equivalent as far as the non-unitary part is concerned: $ e^{-H\delta\tau} $ vs. $e^{H_2 \delta t}$. 
So the same block encoding procedure in the previous section applies, absent of the unitary part $ e^{-iH_1 \delta t}$ and with a sign difference.

\section{Numerical Test}\label{sec:numerics} 

  In this section we put both approaches (imaginary-time and non-Hermitian) to  numerical testing on a quantum simulator, using the exact solutions as a measuring stick.  We are limited to small systems because of the computational demand to run the quantum circuits. 

\subsection{Imaginary time quantum simulation of integrated correlation function}\label{sec:ITEexample} 

In imaginary time, the difference of integrated correlation functions takes the form, 
\begin{equation}
C(\tau ) - C_0(\tau) = \mbox{Tr} \left [   e^{- \hat{H} \tau} - e^{- \hat{H}_0 \tau}   \right ],
\label{Ctau}
\end{equation}
which is real-valued. The imaginary part is completely removed by Wick rotation $\tau = i t$.
We will construct quantum circuits to simulate both the one-qubit Hamiltonian in Eq.(\ref{Honequbit}) and 2-qubit Hamiltonian in Eq.(\ref{Htwoqubit}).

\subsubsection{Single-qubit Hamiltonian system}

The one-qubit Hamiltonian has only two terms.  The second term in Eq.(\ref{Honequbit}) with an identity gate yields a trivial factor, which can be pulled out from the trace: 
\begin{equation}
 \triangle C(\tau )   =  e^{-    \frac{1}{m a^2}   \tau} \left ( e^{-     \frac{V_0}{2 a}   \tau} - 1 \right )  \mbox{Tr} \left [ \left ( e^{ c_X X \delta \tau   }  \right )^N  \right ] , \label{dCtonequbiteq}
\end{equation}
where $\tau = N \delta \tau$, and $c_X =  \frac{1}{m a^2}$.   So we  will  only   compute the non-trivial trace term in Eq.(\ref{dCtonequbiteq}) in quantum simulation to reduce computation cost.

  \begin{figure}[h]
  \frame{\includegraphics[width=0.99\textwidth]{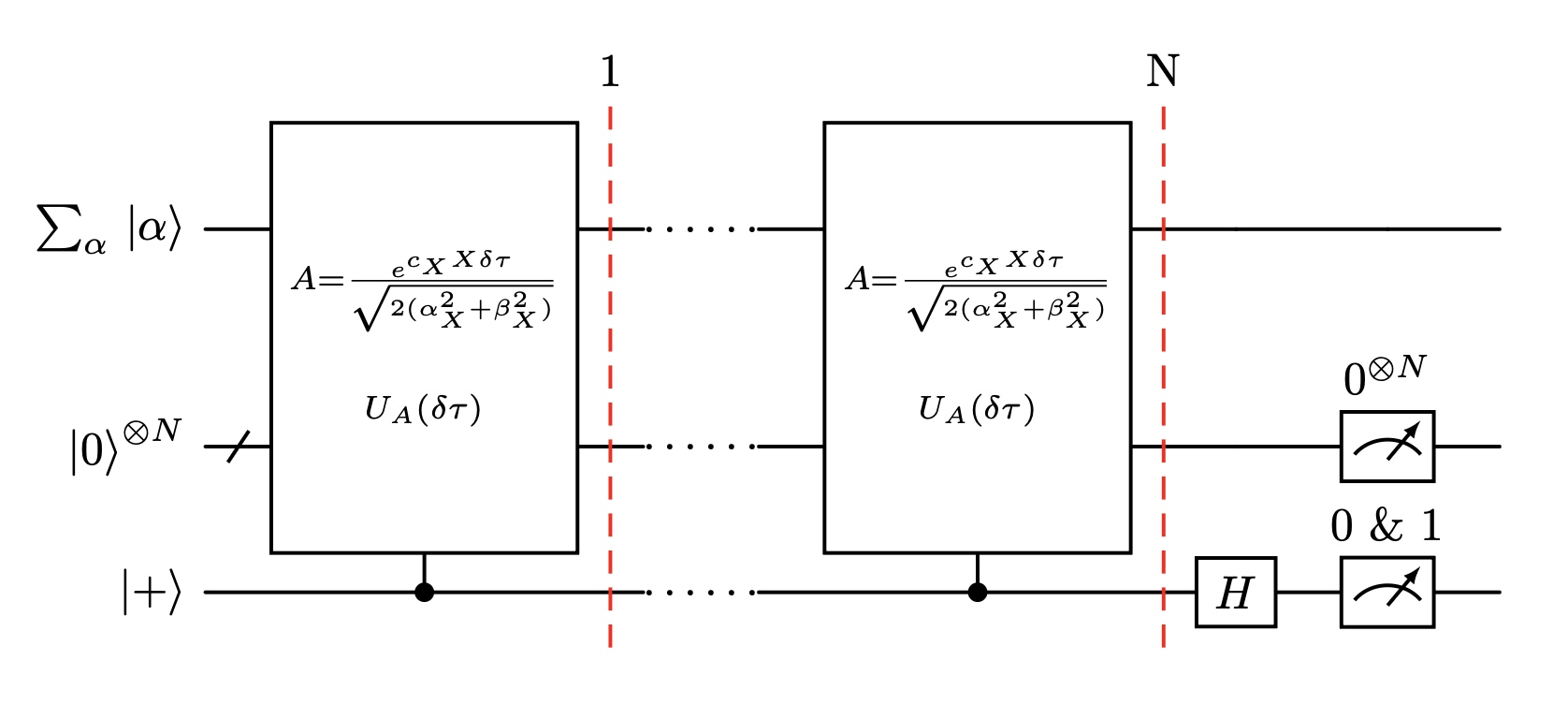}}
    \frame{\includegraphics[width=0.69\textwidth]{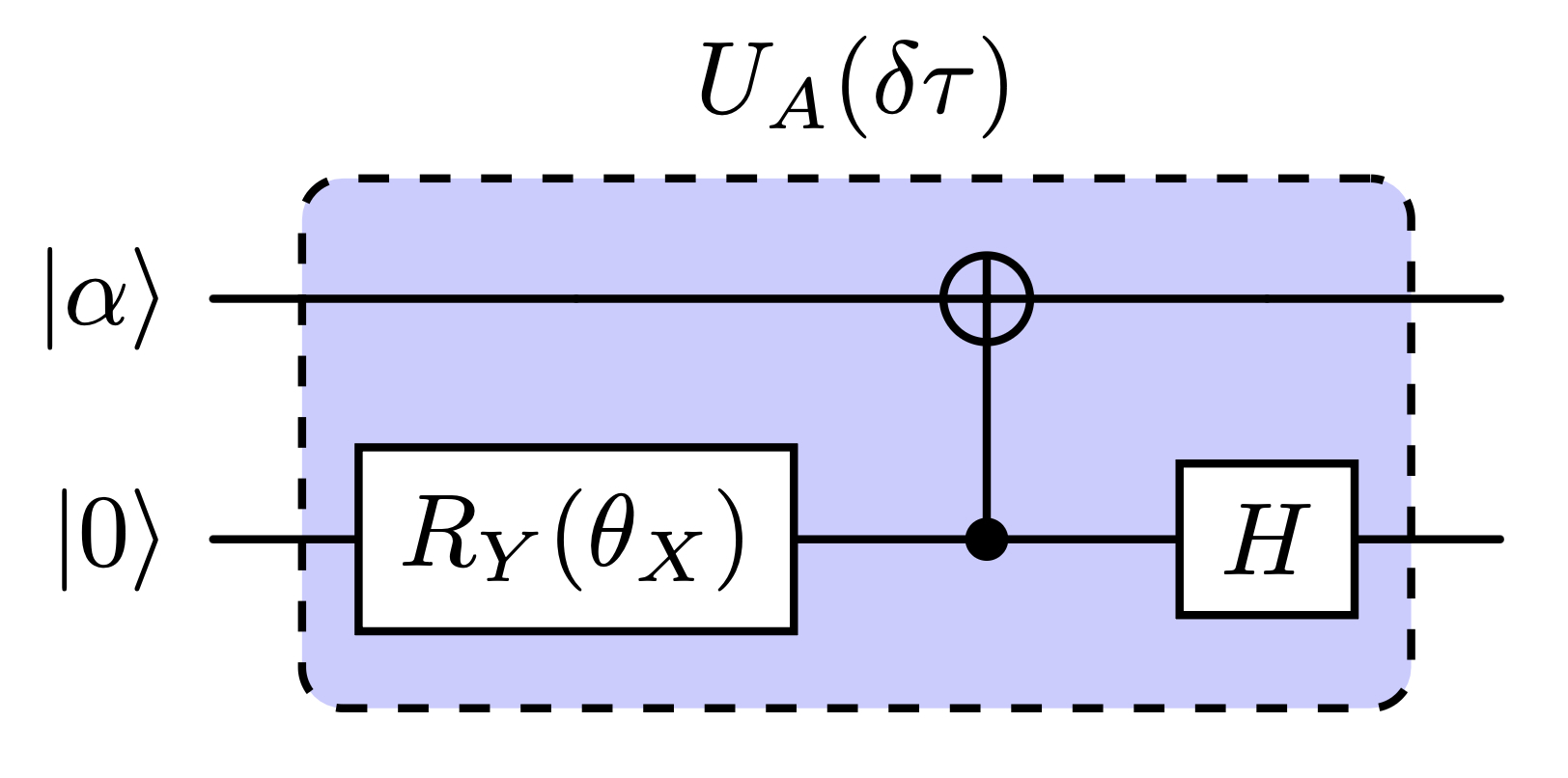}}
\caption{ Quantum circuits for computing  imaginary time evolution of single-qubit system of  Eq.\eqref{traceExpXonequbit} (upper panel), and the block encoding of non-unitary operator $A$ into $U_A (\delta \tau)$ 
(lower panel). }
\label{ITENsteponequbitQC}
\end{figure}

In Fig.~\ref{ITENsteponequbitQC}, we show the block encoding quantum circuits for computing trace of non-unitary operator $e^{ c_X X   \tau   } $ modified by extra normalization factors that are associated to the input $c_X \delta \tau$, 
\begin{equation}
 \sum_\alpha  \langle \alpha |  A^N |  \alpha \rangle \text{ with } 
 A =   \frac{  e^{ c_X  X \delta \tau}  }{\sqrt{2(\alpha_X^2+ \beta_X^2)}}, \label{traceExpXonequbit}
 \end{equation}
 where  
  \begin{equation}
  \alpha_X = \cosh \left ( c_X \delta \tau \right ) ,    \ \  \beta_X = \sinh \left ( c_X \delta \tau \right ) .    \label{defAonequbit}
  \end{equation}
 The $R_Y(\theta_X)$ rotation angle $\theta_X$ is defined in Eq.(\ref{defryangle}).

   \begin{figure}[h]
  \includegraphics[width=0.99\textwidth]{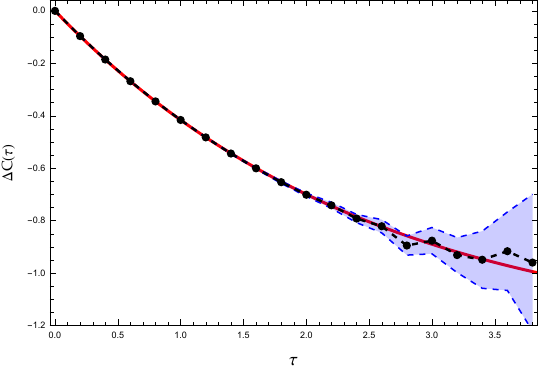}
\caption{Imaginary time simulation of  $\triangle C(\tau)$ for the 1-qubit system in Eq.(\ref{Honequbit}). Red curve is the exact solution. The black dots and blue band represent quantum simulator results for the mean value and two-standard-error band generated from 100 trials, each with  100,000 shots. The model parameters are: $m=1$, $a=4$ and $V_0 =2$. The time step is $\delta \tau=0.2$.}
\label{Fig:dCt1qubit}
\end{figure}

 In this circuit, a single qubit is used for the computation of $\sum_\alpha |  \alpha \rangle $. Each time step requires one ancillary qubit for block encoding $e^{ c_X X  \delta  \tau   } $ into a unitary operation. Thus $N$ time steps require $N$ ancillary qubits for  block encoding  of  the $ \left (e^{ c_X X  \delta  \tau   }  \right )^N$ operator.  Adding another ancillary qubit  for computing  the expectation value $ \langle \alpha |  \left (   e^{ c_X  X \delta \tau}   \right )^N |  \alpha \rangle  $ (Hadamard test), a total of $N+1$ ancillary qubits are needed for quantum simulation of the quantity defined in Eq.(\ref{traceExpXonequbit}). As the number of time steps $N$ is increased, the quantum circuit grows linearly in size both vertically (ancillary qubits) and horizontally (time evolution). Only $0$'s are kept in the measurement of $N$ ancillary qubits for block encoding, and both $0$ and $1$ are kept for the measurement of the additional ancillary qubit for the Hadamard test algorithm.  The  probabilities of $N+1$ ancillary qubits in the state of $| 0 \rangle^{\otimes N}  \otimes | 0 \rangle$ and $| 0 \rangle^{\otimes N}  \otimes | 1 \rangle$  for each   $| \alpha \rangle$ are labeled as  $P_\alpha( 0) $ and $ P_\alpha (1)$, respectively, and the sum of the differences is related to $ \triangle C(\tau )$ by,
 \begin{align}
  \triangle C(\tau ) &  = e^{-    \frac{1}{m a^2}   \tau} \left ( e^{-     \frac{V_0}{2 a}   \tau} - 1 \right ) \left ( \sqrt{2(\alpha_X^2+ \beta_X^2)} \right )^N \nonumber \\
  & \times  \sum_{\alpha} (P_\alpha( 0) - P_\alpha (1))   .
\end{align}
In Fig.~\ref{Fig:dCt1qubit}, we show the comparison of quantum simulator results for  $\triangle C(\tau)$ vs. the exact solution which is obtained by solving Eq.\eqref{Ctau} directly.  The model parameters are chosen as: $m=1$, $a=4$, $V_0=2$, and hence $L=2 a$ according to $a = \frac{L}{2^\Gamma}$. We observe no oscillatory behavior in imaginary-time simulation results, and the agreement with the exact solution extends to over 10 time steps before the signal is lost in statistical fluctuations. As the number of time steps goes up, more statistics are needed in the simulation. One hundred trials of 100,000 shots each are used in this example. The quantum simulator assumes error-free quantum hardware so only statistical errors are  accounted for.

 \begin{widetext}
\subsubsection{Two-qubit Hamiltonian system}
For the two-qubit system Hamiltonian given in Eq.(\ref{Htwoqubit}),  the integrated correlation function with a contact interaction in imaginary time is given by,  
 \begin{equation}
   C(\tau )   =  e^{-    \left (  \frac{1}{m a^2} + \frac{V_0}{4 a} \right )   \tau}  \mbox{Tr} \left [\left(\prod_{j=1}^3 e^{ c_j \hat{h}_j \delta \tau   }\right)^N  \right ], 
\end{equation}
 where all $c$'s and $h$'s are given by,
 \begin{equation}
 c_1 = c_2=\frac{1}{2m a^2} , \;\;c_3 = \frac{V_0}{4 a} ,  
\;\; \hat{h}_1 =  I_2 \otimes  X_1,  \;\; \hat{h}_2 =  X_2 \otimes  X_1,  \;\;\hat{h}_3 =  Z_2 \otimes  Z_1. 
 \label{defAtwoqubitcs} 
 \end{equation}
 In Fig.~\ref{ITENsteptwoqubitQC}, we show  the block encoding quantum circuits for computing,
\begin{equation}
 \sum_\alpha \langle \alpha |  A^N |  \alpha \rangle \text{ with } 
 A= \prod_{j=1}^3  \frac{ e^{ c_j \hat{h}_j \delta \tau   }  }{\sqrt{2(\alpha_j^2+ \beta_j^2)}},
 \label{2qubitA}
 \end{equation}
 where $\alpha_j$ and $\beta_j$ are related to $c_j \delta \tau$ by Eq.(\ref{defAalphabeta}). 
  Two qubits are used for the computation of $\sum_\alpha |  \alpha \rangle $. Three ancillary qubits are needed to block encode  three $\prod_{j=1}^3  e^{ c_j \hat{h}_j \delta \tau } $ non-unitary operators into a unitary operation $U_A$ for a single time step, thus a total of $3N+1$ ancillary qubits are required for $N$ time steps for imaginary time quantum simulation of a two-qubit Hamiltonian system.  
  
   \begin{figure*}[h]
  \frame{\includegraphics[width=0.79\textwidth]{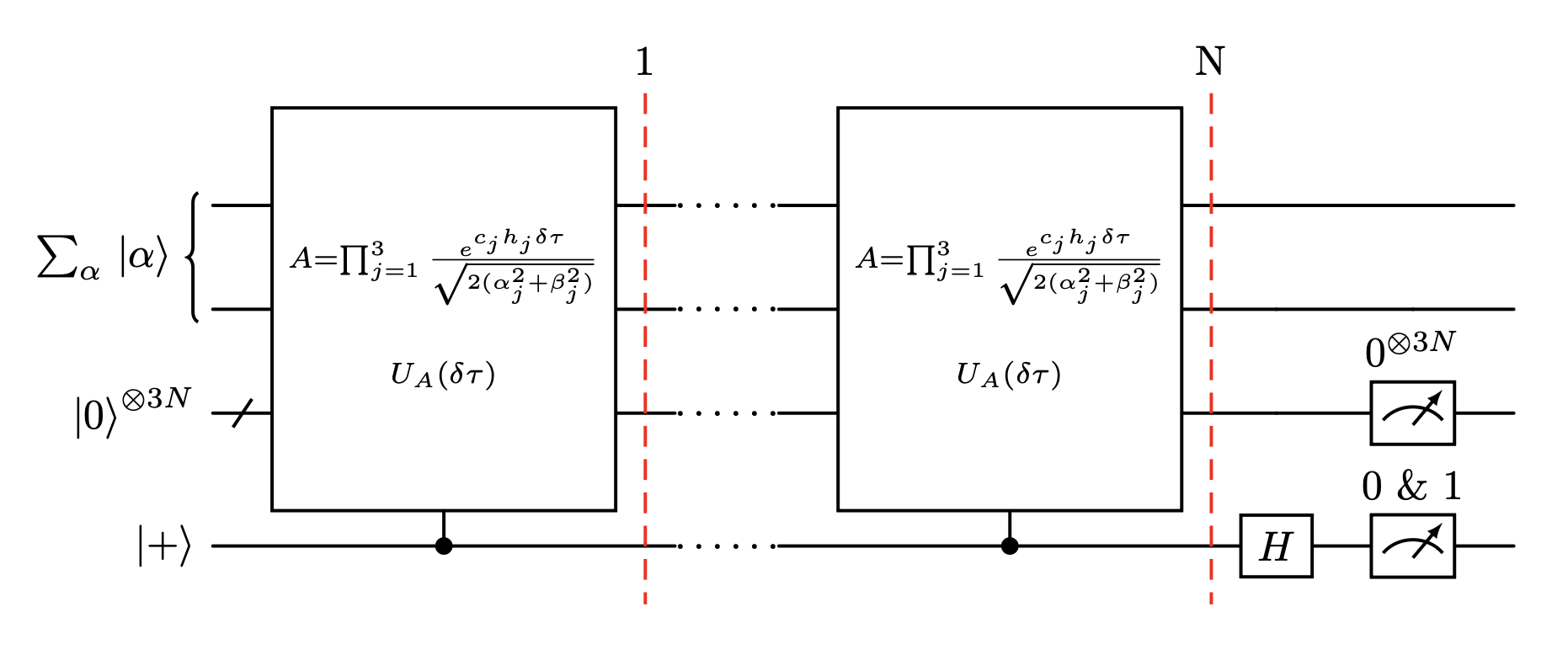}}
    \frame{\includegraphics[width=0.79\textwidth]{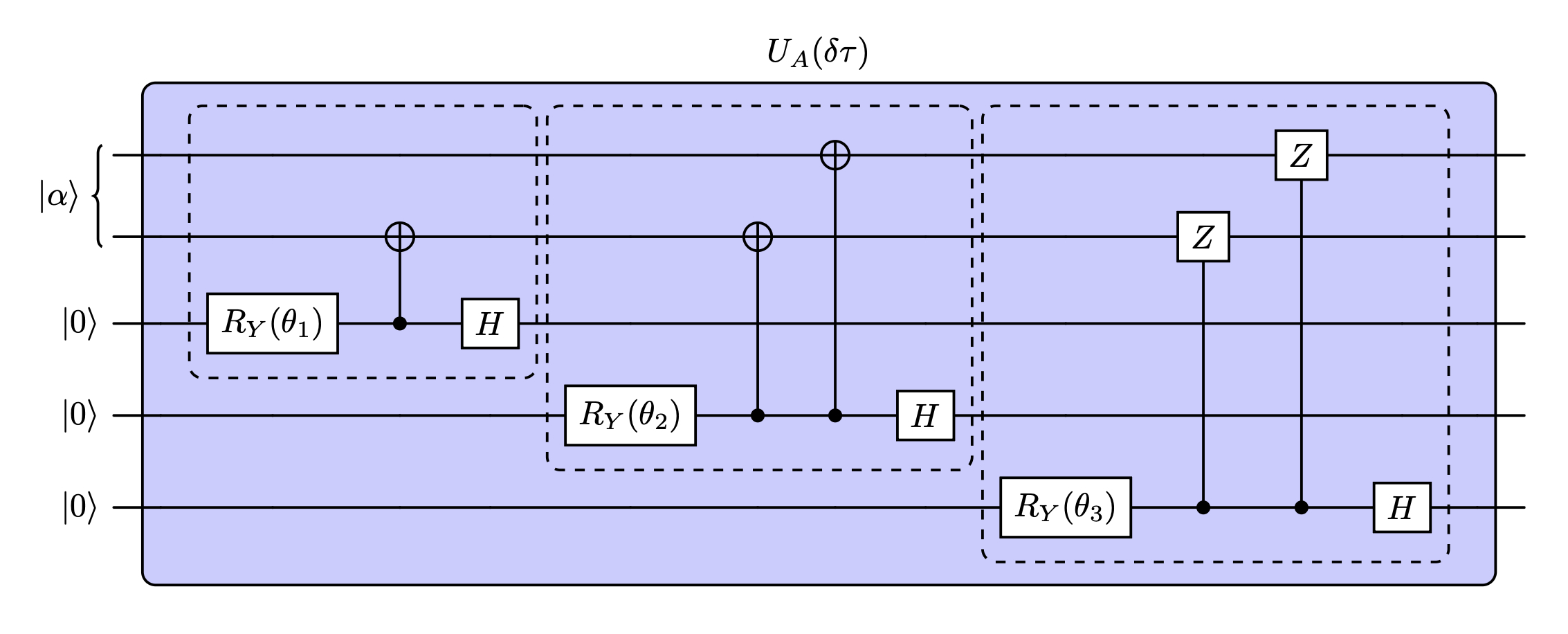}}
\caption{ Quantum circuits   for  computing   imaginary time evolution of two-qubit system of Eq.\eqref{2qubitA} (upper panel), and the block encoding of non-unitary operator $A$ into $U_A (\delta \tau)$ (lower panel). }
\label{ITENsteptwoqubitQC}
\end{figure*} 

 The measurements on the $3 N+1$  ancillary qubits yields,
 \begin{equation}
    C(\tau )   =   e^{-    \left (  \frac{1}{m a^2} + \frac{V_0}{4 a} \right )   \tau}  \left ( \prod_{j=1}^3 \sqrt{2(\alpha_j^2+ \beta_j^2)} \right )^N 
   \times  \sum_{\alpha} (P_\alpha( 0) - P_\alpha (1))   .
  \label{2qubitC}
\end{equation}
The non-interacting $C_0(\tau )$ can be obtained by setting $V_0=0$ (thus $c_3=0$) in the procedure for the interacting case. In Fig.~\ref{Fig:dCt2qubit}, we show the comparison between quantum simulator results and the exact solutions, where the lattice spacing $a$ is adjusted to $4/3$, the rest of model parameters are kept the same  as in the 1-qubit case. The agreement extends to about 5 time steps, then is spoiled by statistical errors. This is expected because it requires $3N+1$ ancillary qubits as opposed to $N+1$ in the 1-qubit case under the same statistics.

   \begin{figure*}[h]
  \includegraphics[width=0.50\textwidth]{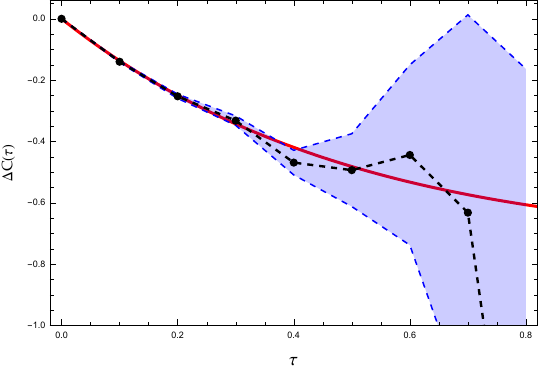}
\caption{Similar to Fig.~\ref{Fig:dCt1qubit}, but for the 2-qubit system Hamiltonian in Eq.(\ref{Htwoqubit}).  The model parameters are: $m=1$, $a=4/3$,   $V_0 =2$  and $\delta \tau=0.1$.}
\label{Fig:dCt2qubit}
\end{figure*}

\subsection{Real time quantum simulation of  integrated correlation function under $L \rightarrow i L$ rotation}\label{sec:nonHermitianexample} 

Under the  $L \rightarrow i L$ rotation,  the single-qubit non-Hermitian Hamiltonian $\hat{H}^{(iL)}$ can be found in Eq.(\ref{HiLonequbit}).  The identity gate term  yields a trivial factor which can be pulled out in the integrated correlation function calculation. The X gate term is Hermitian that yields a unitary operator in quantum simulation and does not require block encoding. Such scenarios have been considered in Ref.\cite{Guo:2026qkx}.  Here we focus on the two-qubit system Hamiltonian given in Eq.(\ref{HiLtwoqubit}) which presents the case study of a non-Hermitian system in real-time quantum simulation.
The integrated correlation function   $C(t )  = \mbox{Tr} \left [ e^{- i \hat{H}^{(iL)} t} \right ]  $ for the interacting Hamiltonian in Eq.(\ref{HiLtwoqubit}) can be expressed as, 
 \begin{equation}
    C(t )   = \mbox{Tr} \left [ \left (  e^{ c_3 \hat{h}_3 \delta t   }  \prod_{j=1}^2 e^{- i c_i \hat{h}_i \delta t }  \right )^N  \right ]   e^{  i \left (  \frac{1}{m a^2} + i \frac{V_0}{4 a} \right )  t}  ,
\end{equation}
where  all $c$'s and $\hat{h}$'s are defined in Eq.(\ref{defAtwoqubitcs}).   The $\prod_{j=1}^2 e^{  - i c_j \hat{h}_j \delta t   }   $ is a unitary operator and does not require block encoding. However, the non-unitary operator $e^{ c_3 \hat{h}_3 \delta t} $ has to be block encoded into a unitary operation that requires one ancillary qubit for each time step. Thus $N$ time steps  require a total of $N+1$ ancillary qubits for quantum simulation of the two-qubit non-Hermitian Hamiltonian $\hat{H}^{(iL)}$ system.

In Fig.~\ref{ITENstepnonHermitianQC}, we give the quantum circuits for computing,
\begin{equation}
 \sum_\alpha \langle \alpha | \left ( A \prod_{j=1}^2 e^{  - i c_j \hat{h}_j \delta t   }   \right )^N |  \alpha \rangle \text{ with } A=\frac{ e^{ c_3 \hat{h}_3 \delta t   }  }{\sqrt{2(\alpha_3^2+ \beta_3^2)}} , 
 \label{2qubitHiL}
 \end{equation}
 where $\alpha_3$ and $\beta_3$ are related to $c_3 \delta \tau$ by definition in Eq.(\ref{defAalphabeta}).  
Two qubits are used for the computation of $\sum_\alpha |  \alpha \rangle $. One ancillary qubit is needed to block encode the $e^{ c_3 \hat{h}_3 \delta t } $ non-unitary operator into a unitary operation $U_A$ for a single time step, thus a total of $N+1$ ancillary qubits are required for $N$ time steps.  Unlike the imaginary-time circuit in Fig.\ref{Fig:dCt2qubit},
the non-Hermitian circuit requires an extra step to handle the unitary operator $\prod_{j=1}^2 e^{  - i c_j \hat{h}_j \delta t   } $. 
 
   \begin{figure*}[t]
  \frame{\includegraphics[width=0.79\textwidth]{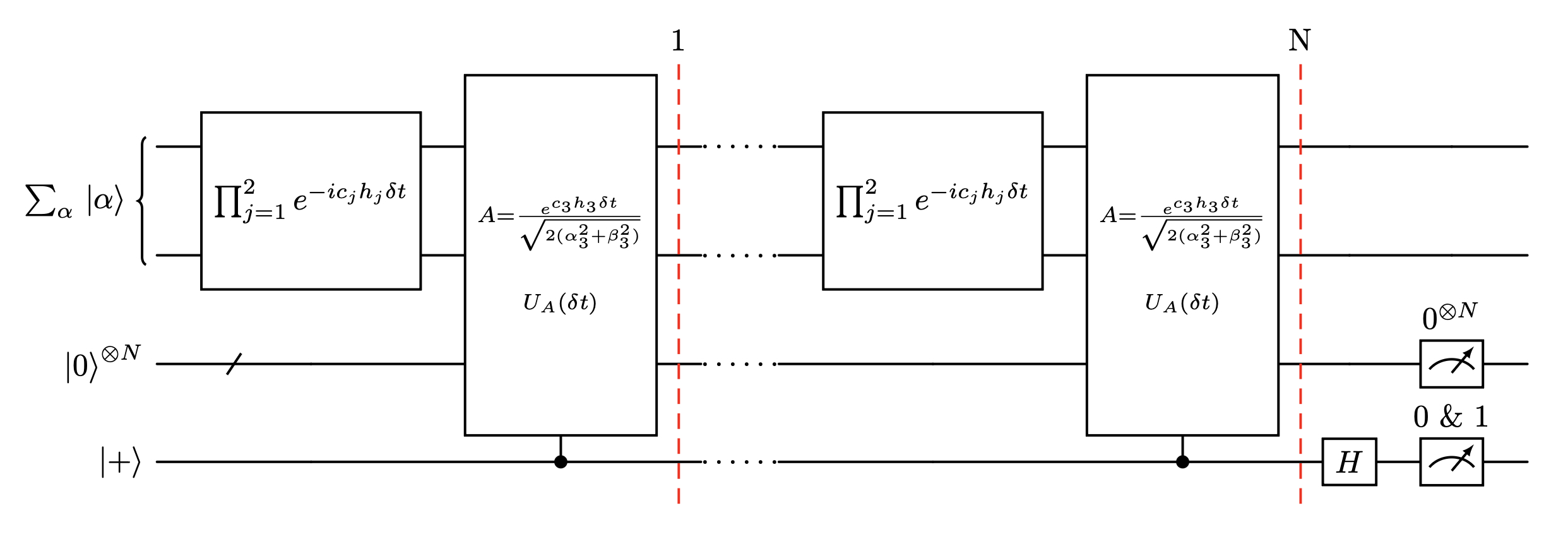}}
    \frame{\includegraphics[width=0.65\textwidth]{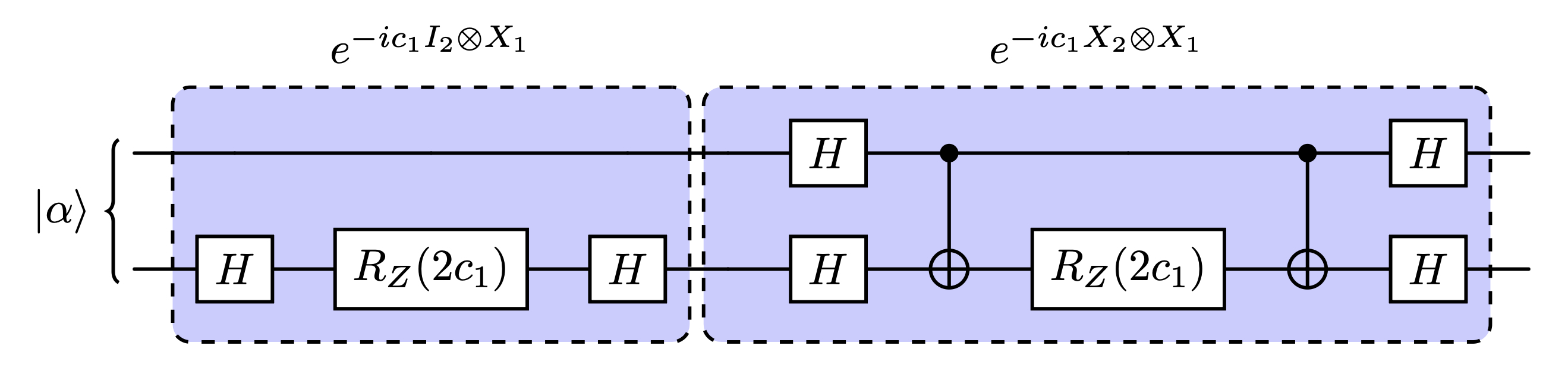}}
      \frame{\includegraphics[width=0.35\textwidth]{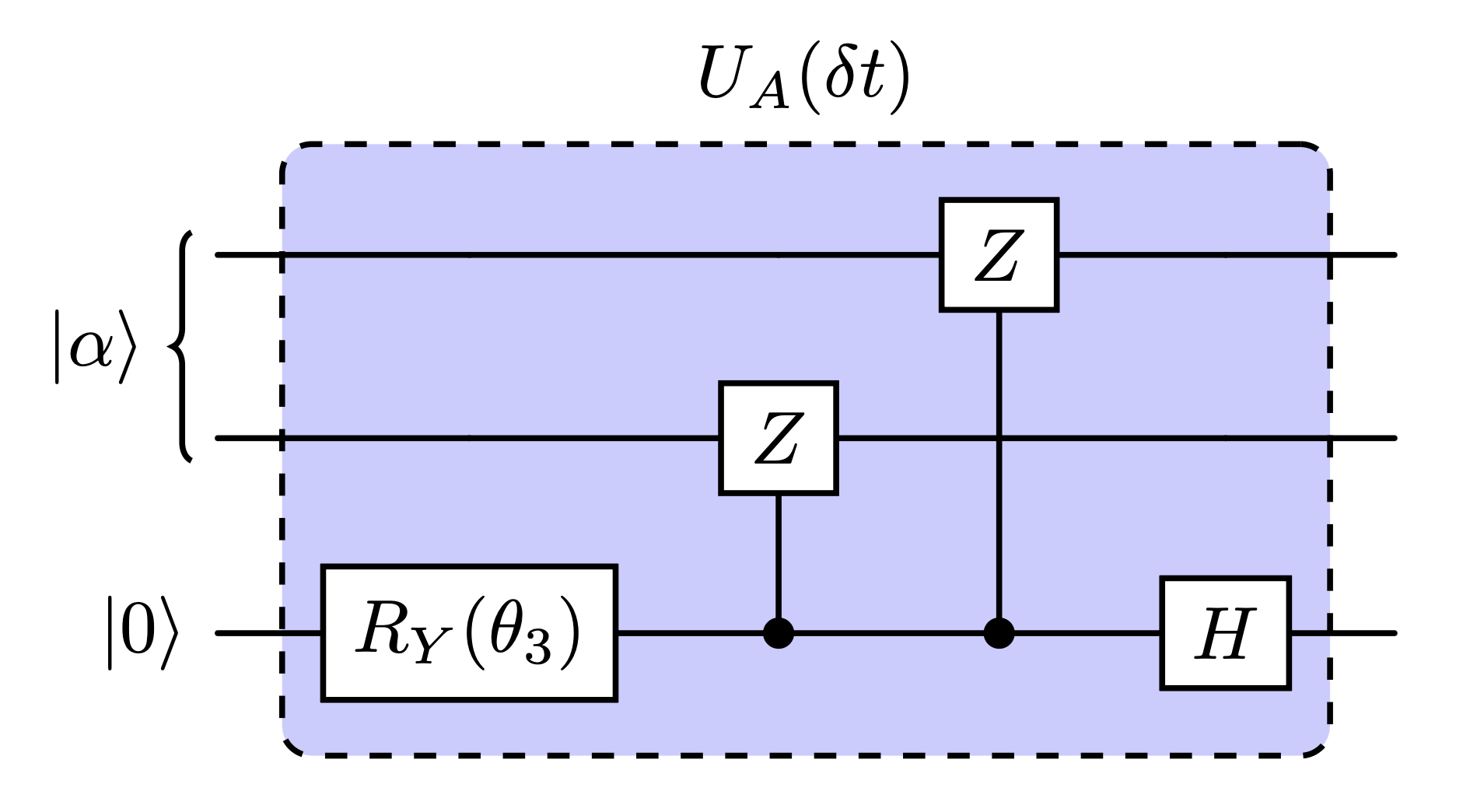}}
\caption{ Quantum circuits   for  computing   real-time evolution of two-qubit non-Hermitian system of Eq.\eqref{2qubitHiL} (upper panel),   unitary operator   $ \prod_{j=1}^2  e^{ - i c_j \hat{h}_j \delta t   }   $ (mid panel), and the block encoding of non-unitary operator $A$ into $U_A (\delta \tau)$ (lower panel).}
\label{ITENstepnonHermitianQC}
\end{figure*} 

  \begin{figure*}[h]
  \includegraphics[width=0.45\textwidth]{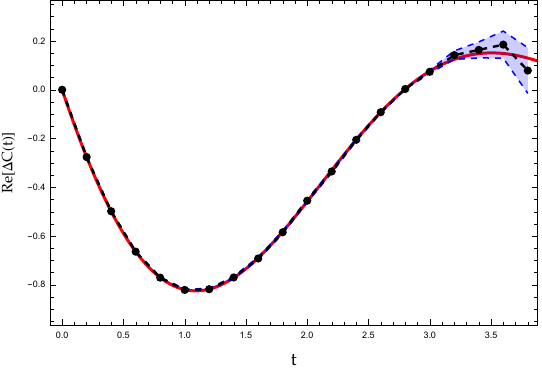}
\includegraphics[width=0.45\textwidth]{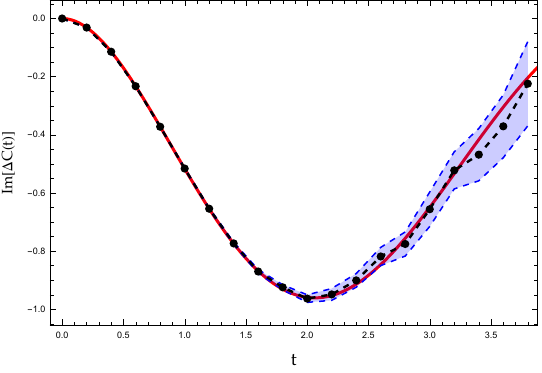}
\caption{Real-time quantum simulation of real (left) and imaginary (right) parts of $\triangle C(t)$ for the two-qubit non-Hermitian system in Eq.(\ref{HiLtwoqubit}).  Red curve is the exact solution. The black dots and blue band represent quantum simulator results for the mean value and two-standard-error band generated from 100 trials, each with 100,000 shots. The model parameters are: $m=1$, $a=4/3$, $V_0 =2$, and $\delta t=0.2$. }
\label{Fig:dCtHiL}
\end{figure*}  

 \end{widetext}
 
  The measurements on the $N+1$  ancillary qubits yields,
 \begin{equation}
    C(t )   =   e^{  i \left (  \frac{1}{m a^2} + i \frac{V_0}{4 a} \right )  t}  \left (   \sqrt{2(\alpha_3^2+ \beta_3^2)} \right )^N   \sum_{\alpha} (P_\alpha( 0) - P_\alpha (1))   .
\end{equation}
 Similarly, the non-interacting $C_0(t)$ can be obtained by setting $V_0=0$ in the procedure for the interacting case.
 In Fig.~\ref{Fig:dCtHiL}, we show the comparison between quantum simulator results and the exact solution using the same model parameters and  one hundred trials of 100,000 shots each. 
The exact result is obtained by solving, 
\begin{equation}
C(t )-C_0(t )  = \mbox{Tr} \left [ e^{- i \hat{H}^{(iL)} t} -e^{- i \hat{H}_0^{(iL)} t}  \right ],  
\end{equation}
 directly for the Hamiltonian in Eq.(\ref{HiLtwoqubit}). 
We see that the agreement extends to around 15 time steps in the real part, and about 10 steps in the imaginary part. This is much better than the 2-qubit case for imaginary-time evolution under the same conditions. 
The reason is simply that the non-Hermitian circuit requires fewer ancillary qubits than in the imaginary-time circuit.

\section{Quantum circuits Optimization for real-time non-Hermitian quantum simulation}\label{sec:optimization}
As demonstrated in Sec.~\ref{sec:numerics}, overall performance of each time step depends heavily on the number of required ancillary qubits to embed the anti-Hermitian part of Hamiltonian into unitary operations. In a rough estimate, the success probability in  both the imaginary-time and the non-Hermitian approaches is proportional to $1/2^n$, where $n$ is the number of ancillary qubits required for block encoding. Though non-Hermitian real-time quantum simulation with $L \rightarrow iL$ rotation approach  reduced the number of required ancillary qubits compared with imaginary-time simulation, fundamentally, it still faces the challenge of inefficiency and practicality in scaling up the size of a quantum system.

In this section, we show  that the quantum circuits in non-Hermitian real-time quantum simulation with $L \rightarrow iL$ rotation approach can be drastically improved and only a single ancillary qubit is required for block encoding in each time step, resulting in a  distinct advantage over the imaginary time approach. With the $ L \rightarrow i L$ rotation, the only anti-Hermitian terms in  Eq.(\ref{HaHbHvcoordinate}) are from the interaction potential term: 
\begin{equation}
V^{(iL)}(x_\alpha)  = \begin{cases} - i \frac{V_0}{2 a }, & \alpha  \in [ \frac{2^\Gamma}{2}-1 ,  \frac{2^\Gamma}{2} ] \\ 0, & \text{otherwise}  \end{cases}   , 
\end{equation}
and $e^{- i \hat{V}^{(iL)} \delta t}$ matrix is a diagonal non-unitary matrix,
\begin{equation}
 e^{- i  \hat{V}^{(iL)}   \delta t}    = \begin{bmatrix} 
1 & \cdots &  0 & 0 &   \cdots & 0  \\
\vdots & \ddots & \vdots & \vdots & \vdots \\
0 &\cdots &  e^{- \frac{V_0 \delta t }{2a}  } &  0 & \cdots  & 0  \\
0 &\cdots &   0 & e^{- \frac{V_0 \delta t }{2 a}  } & \cdots & 0  \\
\vdots & \vdots & \vdots  & \vdots  & \ddots & \vdots \\
 0   &\cdots &    0 & 0 & \cdots & 1
 \end{bmatrix}  .
  \label{eq:expViL}
\end{equation}
It turns out that for a diagonal non-unitary matrix, the insertion of non-unitary terms such as $ e^{- \frac{V_0 \delta t }{2a}  } $ into the certain locations of a diagonal matrix can be accomplished more efficiently than the method used  in Sec.~\ref{sec:blockencode} by  breaking up $\hat{V}^{(i L)}$ into the sum of Pauli strings: $\sum_j c_j \hat{h}_j$ in Eq.(\ref{Hgroups}), and then block encoding each term with one ancillary qubit into unitary operations.

  \begin{figure}[h]
    \frame{\includegraphics[width=0.99\textwidth]{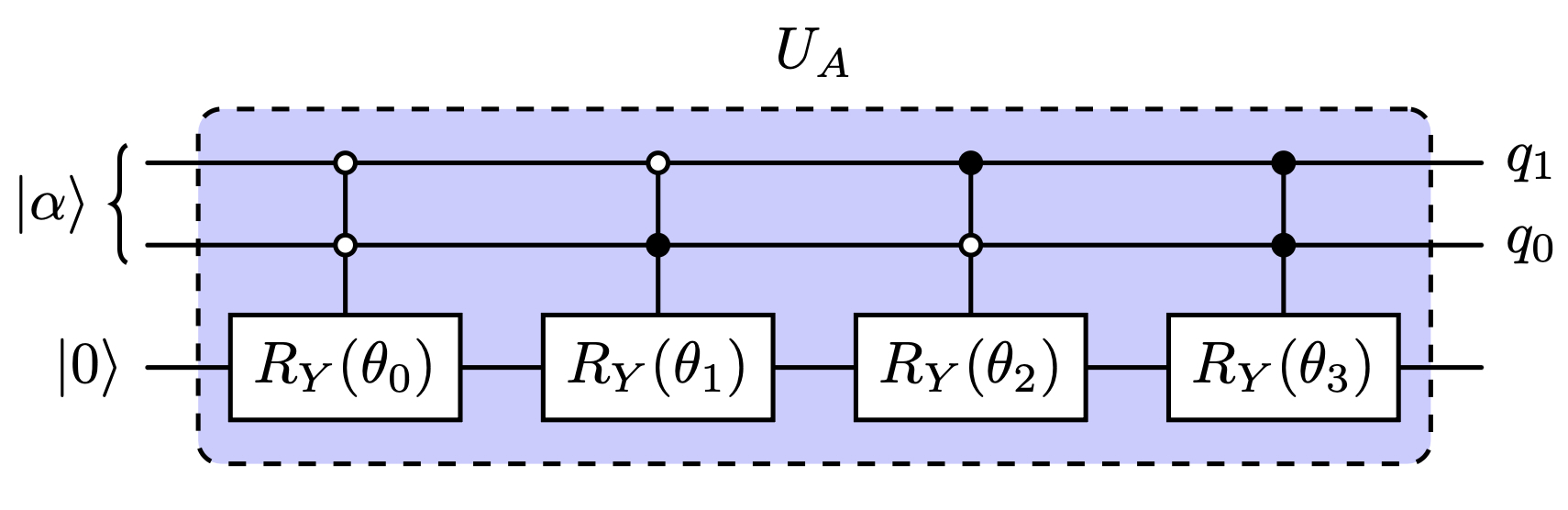}}
\caption{ Quantum circuits for   the block encoding of an diagonal non-unitary operator $A$ in Eq.(\ref{AtwoqubitExample}) into $U_A $. }
\label{fig:UA_twoqubit_example}
\end{figure} 

Using a two-qubit system as a example, the block encoding of a diagonal matrix,
\begin{equation}
A=  \begin{bmatrix} 
e^{- v_0 \delta t  } &   0 & 0 &    0  \\
0 & e^{- v_1 \delta t }   &  & 0  \\
0  &  0 & e^{- v_2 \delta t}   & 0  \\
 0   &   0 & 0  & e^{- v_3 \delta t } 
 \end{bmatrix}, \label{AtwoqubitExample}
\end{equation}
 can be accomplished with a single ancillary qubit and four controlled-$R_Y$ rotation gates in Fig.~\ref{fig:UA_twoqubit_example}. The rotation angles in Fig.~\ref{fig:UA_twoqubit_example} are defined by,
\begin{equation}
\theta_\alpha = 2 \cos^{-1} \left ( e^{- v_\alpha \delta t} \right ), \ \ \ \ \alpha \in [0,3] ,
\end{equation}
where we have assumed that $v_i$'s are all positive and the potential is repulsive.
The controlled-$R_Y (\theta_\alpha)$ rotation gate inserts a factor $e^{- v_\alpha \delta t}$ into the diagonal matrix at location $\alpha$ by using a combination of both states $0$ and $1$ in two qubits that is the exact binary representation of $\alpha$.

If the $v_i$'s are all negative and the potential is attractive, then, we need to block encode the diagonal matrix,
\begin{equation}
B=  \begin{bmatrix} 
e^{  |v_0| \delta t  } &   0 & 0 &    0  \\
0 & e^{  | v_1 | \delta t }   &  & 0  \\
0  &  0 & e^{  |v_2| \delta t}   & 0  \\
 0   &   0 & 0  & e^{ | v_3| \delta t } 
 \end{bmatrix},  
\end{equation}
with the value of all elements greater than one. First of all, by pulling a common factor $e^{\sum_{\alpha =0}^3 |v_\alpha| \delta t}$ out of matrix, we can turn it into a diagonal matrix that has all the value of matrix elements less than one and can be embedded into a unitary operation by using controlled-$R_Y$ rotation gates. Hence the matrix,
\begin{equation}
A = e^{- \sum_{\alpha =0}^3 |v_\alpha| \delta t} B , \label{AtwoqubitExampleAttractive}
\end{equation}
can be block encoded using the quantum circuit in Fig.~\ref{fig:UA_twoqubit_example}, where all the angles are now defined by,
\begin{equation}
\theta_\alpha = 2 \cos^{-1} \left ( e^{- \sum_{\beta \neq \alpha} |v_\beta|  \delta t} \right ), \ \ \ \ \alpha \in [0,3] .
\end{equation}
The common factor $e^{\sum_{\alpha =0}^3 |v_\alpha| \delta t}$ can be installed at the end of quantum simulation at each time step.

In the cases that $v_i$'s are a mixture of negative and positive  values, the diagonal matrix again can be turned into a diagonal matrix with the value of all elements less than one by pulling out a common factor as in the attractive potential case.

 \begin{figure*}[t]
  \frame{\includegraphics[width=0.89\textwidth]{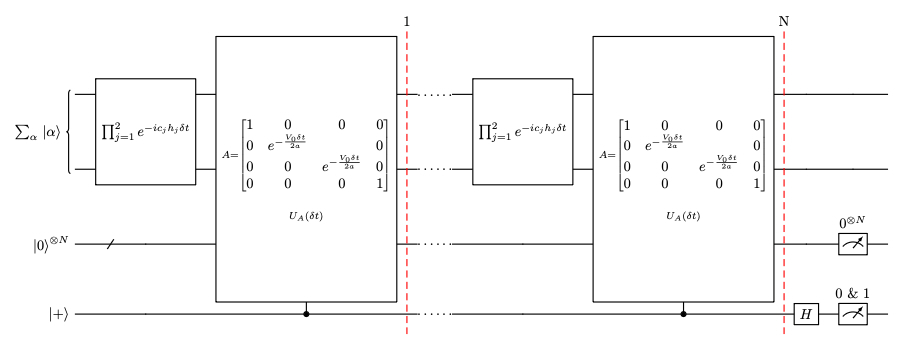}}
     \frame{\includegraphics[width=0.45\textwidth]{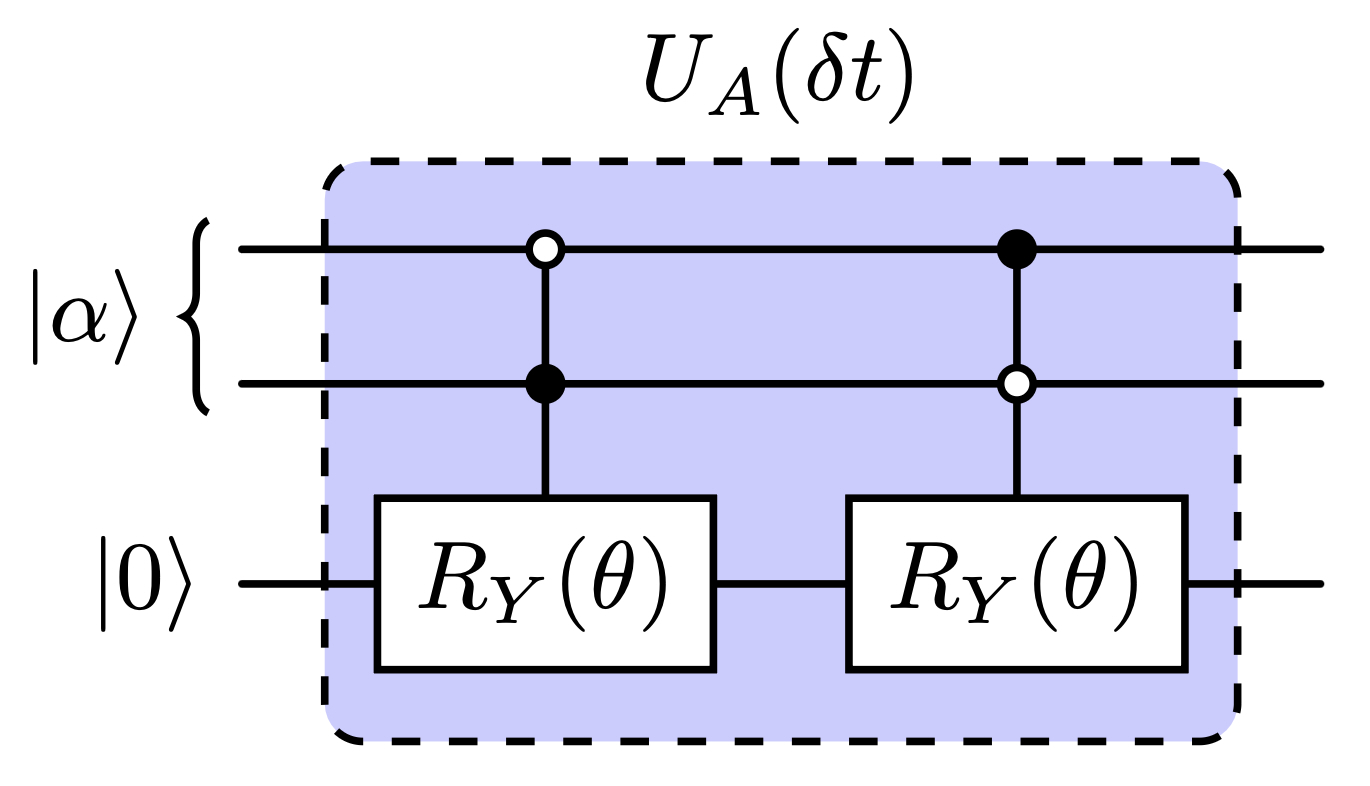}}
\caption{ Quantum circuits   for  computing   real-time evolution of two-qubit non-Hermitian system of Eq.\eqref{ct2qubitHiL} (upper panel),   unitary operator   $ \prod_{j=1}^2  e^{ - i c_j \hat{h}_j \delta t   }   $ are same as the quantum circuit in the mid panel in Fig.~\ref{ITENstepnonHermitianQC}, and the block encoding of non-unitary operator $A=  e^{- i \hat{V}^{(i L)} \delta t} $ into $U_A (\delta \tau)$ (lower panel). The angle $\theta$ in lower panel is defined by $\theta = 2 \cos^{-1} \left ( e^{-  \frac{ V_0 \delta t}{2 a}} \right )$. }
\label{ITENstepnonHermitianQCoptimal}
\end{figure*}

Circling back to the quantum mechanical model with the expression $e^{- i \hat{V}^{(i L)} \delta t}$ given in Eq.(\ref{eq:expViL}),   without losing generality, again using a two qubits system as a simple example,  the integrated correlation function,
 \begin{equation}
    C(t )   = \mbox{Tr} \left [ \left (  e^{- i \hat{V}^{(i L)} \delta t}   \prod_{j=1}^2 e^{- i c_i \hat{h}_i \delta t }  \right )^N  \right ]   e^{  i   \frac{1}{m a^2}    t}   ,\label{ct2qubitHiL}
\end{equation}
can be computed by using the quantum circuit in Fig.~\ref{ITENstepnonHermitianQCoptimal}, where we have assumed that the  potential is repulsive  and  the  $c_{1,2}$ and $\hat{h}_{1,2}$ are defined in Eq.(\ref{defAtwoqubitcs}). By using controlled-$R_Y$ rotation gates to insert $e^{- v_\alpha \delta t}$ into a diagonal matrix $e^{-  \hat{V}^{(i L)}  \delta t}$ as described above, now regardless of the system size, only a single ancillary qubit is required for each time step for block encoding $e^{-  \hat{V}^{(i L)}  \delta t}$ into unitary gates operations. With $N$ steps, a total of $N$ ancillary qubits are needed for block encoding, which is equivalent to the mid-circuit measurement at each time step.

\section{Summary and Outlook}\label{sec:summary} 

In this work,  we explored two proposals to suppress the oscillatory behavior in real-time simulation of quantum systems. We use the same simple case study of integrated correlation scattering formalism in 1D quantum mechanics as in Ref.\cite{Guo:2026qkx}.  One proposal is to simulate a non-Hermitian system in real time, the other is to simulate a Hermitian system in imaginary time. Both proposals lead to the problem of quantum simulation of a non-unitary operator.

We showed that combining two   quantum algorithms (Hadamard test \cite{PhysRevA.75.012328} and block encoding~\cite{PhysRevA.109.052414,Leadbeater_2024,Yi:2025zpa}) can work in both cases. The combined quantum algorithm has a simple form and can be easily implemented. It does not require mid-circuit measurement and the only input is the parameters in the Hamiltonian. 
The algorithm does require that imaginary-time Hamiltonian and the anti-Hermitian part  of the non-Hermitian Hamiltonian (see Eq.\eqref{H1H2}) be expressed in terms of Pauli strings for block encoding. 

In terms of computational demand, 
the size and length of the quantum circuits grow linearly as the number of time steps increases in both approaches. Even for small-size Hamiltonian systems, running tests on local quantum simulators becomes a computational challenge after 10 to 20 steps. 
We constructed 1-qubit and 2-qubit Hamiltonian systems to numerically test the ideas against exact solutions. In both cases, we observe  good agreement with the exact solutions to reasonably long times before the signal is overwhelmed by statistical fluctuations.

The non-Hermitian simulation does have  a computational advantage over the imaginary-time one. Comparing at the 2-qubit Hamiltonian level, there are three non-unitary terms in imaginary-time Hermitian simulation that need to be block encoded, versus only one term in real-time non-Hermitian simulation. Block encoding of each non-unitary term introduces one extra ancillary qubit to make it into a unitary operation. Hence, the cost of imaginary-time quantum simulation is three times higher than that of non-Hermitian one.

The success probability in  both  the imaginary-time and the non-Hermitian approaches is proportional to $1/2^n$, where $n$ is the number of ancillary qubits required for block encoding. This makes the simulations appear costly and impractical at scale. Fortunately,   the quantum circuits for real time non-Hermitian quantum simulations can be optimized so that  only a single ancillary qubit is required for block encoding non-unitary diagonal  $e^{- i \hat{V}^{(i L)} \delta t}$  matrix for each time step. Hence even without amplitude amplification algorithms,  the correlation function can be computed fairly accurately up to at least ten steps with 100,000 shots.  The  $ L \rightarrow i L$ rotation approach definitely boosts our confidence for the practicality  of      quantum simulation at scale.  How the $ L \rightarrow i L$ rotation works in  other field theory models, such as the $\phi^4$ theory, and whether the quantum simulation of a non-Hermitian system is practical in general, are not clear at this stage and deserve further study.

Looking forward, 
although the work performed here is limited to a simple quantum mechanical model and small systems, we expect the imaginary-time and non-Hermitian approaches to be helpful in simulation of other quantum systems on quantum computers.
Finally, it  should be mentioned that the computational challenge mentioned above is due to running quantum simulators on classical computers.  If the circuits are executed on fault-free quantum hardware,   quantum   advantage is expected.

\acknowledgments
This research is supported by the U.S. National Science Foundation under grant PHY-2418937 (P.G.) and PHY-2531653 (P.G.), and by the U.S. Department of Energy under grant  DE-FG02-95ER40907 (F.L.),   DE-AC02-06CH11357 (Y.Z.) and DOE Early Career Award through Contract No.~DE-SCL0000017 (Y.Z.).

\appendix

\section{Derivation of the Pauli Basis Expression for \(\hat{H}_b\)}\label{sec:appendix-A}

In this appendix we derive the Pauli basis representation of
\begin{equation}
B_\Gamma = \sum_{\alpha=0}^{2^{\Gamma-1}-1}
\left(
|2\alpha+1\rangle\langle 2\alpha+2|
+
|2\alpha+2\rangle\langle 2\alpha+1|
\right)
\label{defBn_app}
\end{equation}
where we identify \( |2^\Gamma\rangle\equiv |0\rangle\), so that \(\hat{H}_b\) is obtained by multiplying by \(\dfrac{-1}{2ma^2}\). Here the computational basis is ordered in bitwise representation as
\begin{equation}
|\alpha\rangle = |b_\Gamma\cdots b_1\rangle,
\qquad
\alpha=\sum_{j=1}^\Gamma 2^{j-1}b_j,
\end{equation}
with \(b_1\) the least significant bit.

The operator in Eq.~(\ref{defBn_app}) swaps the basis-state pairs
\begin{equation}
0 \leftrightarrow 2^\Gamma-1,
\;\;
1 \leftrightarrow 2,
\;\; \dots \;\;
2^\Gamma-3 \leftrightarrow 2^\Gamma-2
\label{Bn_pairs}
\end{equation}

If we split the basis according to the most significant bit by setting \(|b,\mu\rangle \equiv |b\rangle\otimes |\mu\rangle\) where \(b\in\{0,1\}\)
and \(|\mu\rangle\) denotes the last \(\Gamma-1\) qubits, then, away from \(\mu=0\) and \(\mu=2^{\Gamma-1}-1\), the pairings in Eq.~(\ref{Bn_pairs}) are identical in the \(b=0\) and \(b=1\) halves for the respective \((\Gamma-1)\)-qubit actions on \(|\mu\rangle\). The only changes occur at the corners with:
\begin{align*}
|0\rangle\otimes |0^{\Gamma-1}\rangle &\leftrightarrow |1\rangle\otimes |1^{\Gamma-1}\rangle \\
|0\rangle\otimes |1^{\Gamma-1}\rangle &\leftrightarrow |1\rangle\otimes |0^{\Gamma-1}\rangle
\end{align*}
This means we can recover the operator \(B_\Gamma\) by starting from two copies of the \((\Gamma-1)\)-qubit operator \(B_{\Gamma-1}\) and replacing the two internal corner swaps by two half swaps.

In terms of a matrix expression, we can write this by defining
\(W_n
=
|0^n\rangle\langle 1^n|
+
|1^n\rangle\langle 0^n|\)
and writing
\begin{equation}
B_n=\begin{pmatrix}
B_{n-1}-W_{n-1} & W_{n-1} \\
W_{n-1} & B_{n-1}-W_{n-1}
\end{pmatrix}
\end{equation}
or equivalently, \(B_n = I_n\otimes B_{n-1} + (X_n-I_n)\otimes W_{n-1}\) with \(B_1=X_1\).

For one qubit,
\[
|0\rangle\langle 1|=\frac{X+iY}{2},
\qquad
|1\rangle\langle 0|=\frac{X-iY}{2}
\]
And so
\begin{align}
W_n
&=
|0^n\rangle\langle 1^n|+|1^n\rangle\langle 0^n| \nonumber \\
&=
\frac{1}{2^n}
\bigl[
(X_n+iY_n)\otimes\cdots\otimes (X_1+iY_1)\\
&\qquad\quad+
(X_n-iY_n)\otimes\cdots\otimes (X_1-iY_1)
\bigr]
\label{Wn_start}
\end{align}

For each subset \(S\subseteq\{1,\dots,n\}\), if we define
\begin{equation}
\begin{split}
&P_{k-1}(S)=M_{k-1}(S)\otimes\cdots\otimes M_1(S)\\
&M_j(S)=
\begin{cases}
Y_j, & j\in S,\\
X_j, & j\notin S.
\end{cases}
\end{split}
\label{P_def}
\end{equation}
Then choosing \(Y_j\) exactly on the subset \(S\) contributes coefficient \(i^{|S|}\) from the first product in Eq.~(\ref{Wn_start}) and \((-i)^{|S|}\) from the second after expanding these binomials. Hence the coefficient of \(P_n(S)\) in \(W_n\) is \(\frac{1}{2^n}\left(i^{|S|}+(-i)^{|S|}\right)\).
If \(|S|\) is odd, this vanishes, whereas if \(|S|=2r\) is even, then \(i^{S|}+(-i)^{|S|}=2(-1)^r=2(-1)^{\frac{S}{2}}\). Together with the recursion for \(B_n\), this iterates to the expression Eq.~(\ref{hatB_main}) for \(\hat{H}_b\).
 
\bibliography{ALL-REF.bib}

\end{document}